\documentclass[twocolumn,twocolappendix]{aastex631}
\usepackage{graphicx}
\usepackage{subcaption}
\usepackage{natbib}
\usepackage{booktabs} 
\usepackage{hyperref}
\usepackage{array}
\usepackage{multirow}
\usepackage{amsmath}

\usepackage{float}
\newcommand{\NSide}{$N_{Side}$}

\begin{document}

\renewcommand\linenumberfont{\normalfont\tiny\sffamily}

\let\oldfigure\figure
\let\endoldfigure\endfigure
\renewenvironment{figure}
  {\nolinenumbers\oldfigure}
  {\endoldfigure}

\let\oldtable\table
\let\endoldtable\endtable
\renewenvironment{table}
  {\nolinenumbers\oldtable}
  {\endoldtable}

\let\oldfigurestar\figure*
\let\endoldfigurestar\endfigure*
\renewenvironment{figure*}
  {\nolinenumbers\oldfigurestar}
  {\endoldfigurestar}

\let\oldtablestar\table*
\let\endoldtablestar\endtable*
\renewenvironment{table*}
  {\nolinenumbers\oldtablestar}
  {\endoldtablestar}
\title{STARFIRE-2: Can we detect the global redshifted 21-cm signal from the cosmic dawn in Earth orbit?}

\author{Yogen Pranesh}
\affiliation{Department of Physics, University of Rome Tor Vergata, 00133 Rome, Italy}
\affiliation{Raman Research Institute, C.~V.~Raman Avenue, Sadashivanagar, Bangalore 560080, India}

\author{Mayuri Sathyanarayana Rao}
\affiliation{Raman Research Institute, C.~V.~Raman Avenue, Sadashivanagar, Bangalore 560080, India}

\author{Saurabh Singh}
\affiliation{Raman Research Institute, C.~V.~Raman Avenue, Sadashivanagar, Bangalore 560080, India}



\begin{abstract}
Detecting the redshifted global 21-cm signal from the cosmic dawn (CD) remains a major challenge due to strong terrestrial Radio Frequency Interference (RFI), particularly dominated by Frequency Modulation (FM) transmissions in the 88-110 MHz range. While observations from the radio-quiet lunar farside are ideal, Earth orbit offers an intermediate and simpler alternative that may mitigate several limitations of ground-based experiments. 

We assess the feasibility of detecting the global 21-cm signal from Earth orbit by quantifying FM-based RFI at different altitudes and orbital configurations. We present STARFIRE-2 (Simulation of TerrestriAl Radio Frequency Interference in oRbits around Earth -- 2), an algorithm that estimates FM transmitter--based RFI intercepted by radiometers in orbit. The model constructs a global FM transmitter database and compensates for incomplete data using statistical methods. Using PRATUSH as the reference experiment, we simulate a range of orbital scenarios to identify configurations that minimize RFI and optimize sensitivity for global 21-cm detection. The algorithm can also be adapted for other experiments. 

Simulations indicate that conducting such an experiment from Earth orbit is feasible for a thermal noise limited instrument placed in a low-Earth, near-polar orbit. Mock sky observations further demonstrate that most theoretically plausible cosmic dawn 21-cm signals can be recovered with high confidence under these optimized orbital conditions.
\end{abstract}



\section{Introduction} \label{sec:intro}
It has been long predicted that the astrophysics and cosmology of the dark-ages, cosmic dawn (CD) and epoch of reionization (EoR) can be studied using the redshifted 21-cm signal from hydrogen \citep{21_in21} either as a monopole global signal or as multi-point correlation functions such as power spectrum and beyond. Throughout this paper, we use the terms monopole, sky-averaged signal, and global signal interchangeably. Since then several ground-based experiments have searched for the faint global signal \citep{Singh01,Singh02,Singh03,Singh04,Bowman02,Philip01,Bernardi01, Price01,REACH}, still a confident detection remains elusive.  Ground-based experiments aiming to detect the redshifted global 21-cm signal from CD/EoR encounter numerous challenges. For instance chromatic effects induced in the measured sky spectrum from ionospheric refraction, reflection and absorption can limit foreground separation and hence signal detection \citep{vedantham_chromatic_2014,datta_effects_2016,shen_quantifying_2021}. Also, antenna properties depend critically on the operating environment including, but not limited to, ground-plane effects, dielectric constant of the near-field, which vary in ways that are complex and hence difficult to model \citep{agrawal_direction-dependent_2024, bradley_ground_2019, Antenna_char}. Finally, terrestrial Radio Frequency Interference (RFI) presents a significant challenge for CD/EoR experiments (\citet{RFI_bad}, \citet{SSINS}). This issue is particularly problematic in the FM band (88 to 110 MHz), which has signigicant overlap with the frequency range of CD observations. Several experiments have been proposed to measure the low frequency radio sky spectrum in the pristine radio environment of the lunar farside \citep{DARE,DAPPER,DSL}. PRATUSH – Probing ReionizATion of the Universe using Signal from Hydrogen, is one such proposed experiment from India that aims to detect the CD/EoR signal from the lunar farside \citep{PRATUSH}.
 
Before heading to the moon, one can study the feasibility of conducting this experiment in free space, far away from the surface of the Earth, to alleviate challenges posed by the ionosphere and medium / environment induced antenna chromaticity. An experiment in Earth orbit as a precursor to a lunar orbiter provides an intermediate option with lower logistic and economic complexity, with more opportunities for launch and technological readiness  and sensitivity. Whether such an experiment can generate competitive results depends critically on the RFI environment in Earth orbit, thus motivating this work.  
Since an experiment in Earth orbit will have a Field Of View - FOV - (of Earth) proportional to its altitude, a space-based radio telescope will inevitably capture more number of RFI sources within its beam than perhaps one observing at a radio quiet location on Earth. However, it will certainly be devoid of the other two challenges faced by ground-based experiments, namely ionospheric effects and antenna-coupling to the surrounding medium. In a previous paper we presented STARFIRE- Simulation of TerrestriAl Radio Frequency Interference in oRbits around Earth, an algorithm to estimate FM transmitter based RFI at different Earth orbits \citep{STARFIRE}. While the algorithm proved effective, the user-provided FM transmitter information in the database used in STARFIRE was sparse and hence the final results were of limited application. Recognizing the need for more complete data, we propose a method for building a statistically complete FM transmitter database and conducting forecast studies for PRATUSH in Earth orbit, henceforth referred to as PRATUSH-I. We take this effort one step further, to quantify the feasibility of conducting a CD/EoR experiment in Earth orbit by examining the signal detection prospects predicted by standard models. \par
In Section \ref{sec: FMDB}, we build a complete FM transmitter database and account for incomplete information by statistically modeling various parameters and generating synthetic data, followed by Section \ref{sec: orbit sims}, where we conduct various orbit simulations to estimate the amount of RFI intercepted by PRATUSH-I throughout its orbital lifetime. The radio quiet zones in the orbit would thereby dictate the feasibility of conducting such an experiment in Earth orbit. Section \ref{sec: sky obs} presents simulations of sky observations by PRATUSH at different points in its orbit path and quantify the sensitivity of the radiometer to various standard CD models. We discuss sources of uncertainty in section \ref{sec: Discussion} and present conclusions in section \ref{sec: Conclusions}.

\section{Building a complete FM transmitter database}\label{sec: FMDB}
The STARFIRE algorithm uses as its input an FM transmitter database that includes each FM transmitter's geodetic coordinates, its frequency of transmission, and its Effective Isotropic Radiated Power (EIRP). Folding in instrument properties such as antenna beam, frequency span and resolution, it generates an RFI spectral cube for user-defined satellite altitudes. The algorithm uses the HEALPix pixel scheme \citep{HEALPix} in Python using the `Healpy'  library \citep{Zonca2019}. Accurate RFI estimation for a receiver in Earth orbit using STARFIRE critically depends on the completeness of the input RFI source database, which in this case is the FM transmitter database. Such a database has been populated using FMLIST\footnote{\href{https://www.fmlist.org/}{FMLIST.org}}, a publicly available worldwide FM transmitter database that sources its data from government websites and HAM radio operators. This open-access country-wise FM transmitter information was processed to match the specifications required by the STARFIRE algorithm and added to the database.\par


\subsection{Incomplete datasets for select countries}\label{subsec: Incomplete datasets}
FM transmitter information on the FMLIST interface is displayed in ascending order of frequency of transmission. One can access information for up to the first 5000 transmitters per country in the free version of FMLIST. The interface also displays the total number of FM transmitters in the country. Thus, transmitter datasets of countries with more than five thousand transmitters (See Table \ref{tab:missing Tx}) are incomplete in the free version of FMLIST. This can result in underestimating the RFI by the STARFIRE algorithm. We build a statistical model of FM transmitters to keep the database flexible and future-proof (such as FM transmitters coming up / going down in the future).  This statistical approach not only accounts for the remaining data but also adds scalability to the STARFIRE input database, providing flexibility to replace FMLIST with other transmitter datasets if needed. Here we note that using a statistical model to account for missing transmitters instead of exact transmitter data does not alter the output of the STARFIRE algorithm significantly. It is unlikely to convert a radio-quiet location or channel into a radio-loud one (or vice versa), because any additional statistically generated transmitters would primarily appear in countries that already exceed 5000 registered transmitters—regions that are inherently high-RFI to begin with. To account for the `missing' transmitters, we populate synthetic transmitters by generating the location, frequency, and EIRP parameters per transmitter using statistical methods detailed below. We note that though the USA has more than 5000 FM transmitters, we have full data available from FCC\footnote{\href{https://www.fcc.gov/media/radio/fm-query}{FCC.gov}} (Federal Communications Commission). This dataset was used as a reference dataset for testing the effectiveness of our statistical methods used to build a complete database.

\begin{table}[h]
\caption{Number of missing transmitters for countries with incomplete databases.}
\label{tab:missing Tx}
\centering
\begin{tabular}{lc}
\hline
Country & Number of missing Tx \\
\hline
Russia     & 10\,508 \\
China      & 4\,333  \\
Argentina  & 2\,092  \\ 
Brazil     & 4\,201  \\
Spain      & 1\,340  \\
France     & 4\,406  \\
\hline
\end{tabular}
\end{table}

\begin{figure}[ht]
  \centering
  \begin{subfigure}{0.45\linewidth} 
    \centering
    \includegraphics[width=\linewidth]{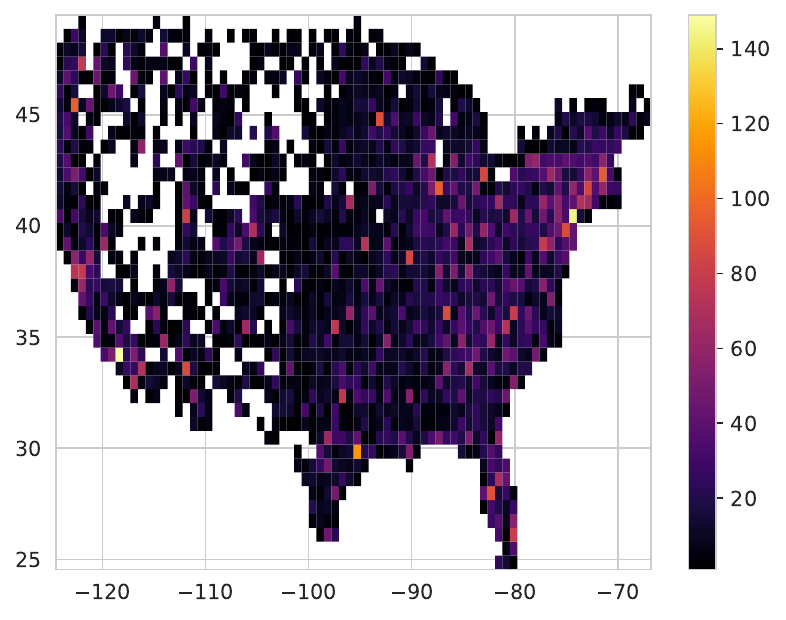}
    \caption{Spatial distribution of FM transmitters in USA with complete database}
  \end{subfigure}
  \hfill 
  \begin{subfigure}{0.45\linewidth} 
    \centering
    \includegraphics[width=\linewidth]{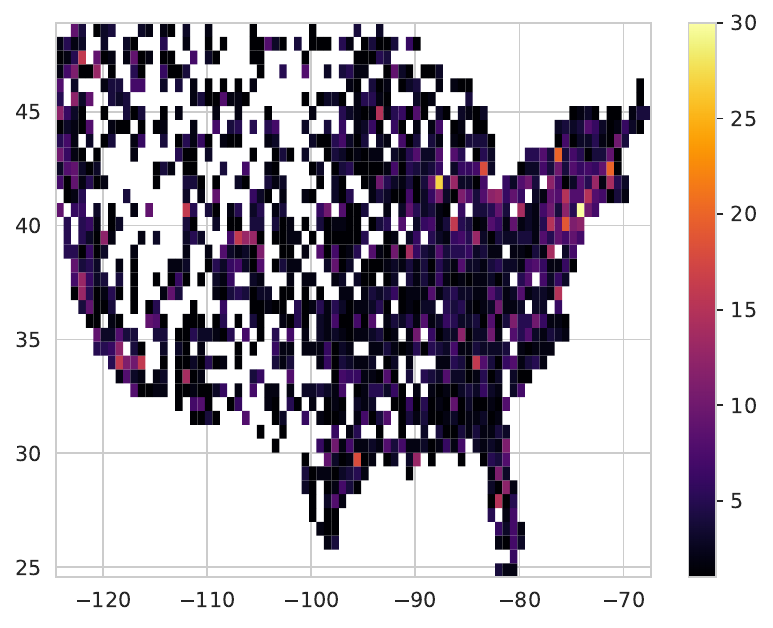} 
    \caption{Spatial distribution of FM transmitters in USA with incomplete database}
  \end{subfigure}
  \caption{Spatial transmitter distribution comparison}
  \label{fig:dist_comp}
\end{figure}

\subsubsection{Modeling FM transmitter location parameter}\label{subsubsec: location}
The area of the geometric footprint of a satellite is given by \begin{equation}
    A = 2\pi R_e^2 \left(1 - \frac{R_e}{R_e + h}\right),
\end{equation} 
where $R_e$ is the radius of the Earth, and $h$ is the altitude of the satellite above the mean sea level. For a satellite in Low-Earth-Orbit ($\sim 410$ km altitude), the geometric footprint, or field of view, is expected to cover 15 million square kilometers, making its footprint the smallest among all Earth orbits (Low Earth orbits, medium Earth orbits, and geostationary Earth orbits). This sets the requirement of the finest spatial resolution of the pixel when placing synthetic transmitters in the statistical model. As a result, the accuracy of synthetic transmitter locations, when compared to their real counterparts, is not highly constrained.

Since the location parameter of a transmitter is bivariate (latitude and longitude), a normalized bivariate probability distribution function was created using a two-dimensional histogram of transmitter latitude and longitude using the \texttt{histogramdd} function of the NumPy python library \citep{harris2020array}. We assume that the 2-D distribution, created out of 5000 points, is a good representation of the full distribution of the transmitters in the country. A transmitter location corresponds to a coordinate pair within the geo-political boundary of the country it is located in. 
Hence sampling from the spatial distribution must be done using ordered-pairs which also follow the geometry of the country corresponding to its geo-political boundary. A HEALPix pixel located within a country represents an ordered-pair, as it corresponds to a specific longitude and latitude value.
\par
The accuracy of generating locations of synthetic FM transmitters using the prior spatial probability distribution functions strongly depends on the bin size. Since the bin width dictates the angular resolution of sampling for the probability distribution, a large bin width would not capture the intrinsic spatial variation in the distribution. In contrast, a very small bin width would tend to over-fit the distribution by capturing random fluctuations \citep{bin_fluctuation}.  
The optimal bin width was chosen by minimizing the `cost function' across various selections of bin widths as directed by \citet{shimazaki2007method}. The cost function is defined as 
\begin{equation}
    C(\Delta)= \frac{2k-v}{\Delta^2},
\end{equation} 
where $k$ and $v$ are the mean and standard deviation of the number of variables per bin and $\Delta$ is the selected bin width for the histogram.
It is feasible to generate synthetic transmitter locations using the incomplete dataset transmitters' spatial distribution as a 2-D probability distribution function. 
\par
One can generate the (latitude,longitude) ordered pairs of coordinates within a country at a chosen angular resolution by obtaining the `shapefile' of the country. A shapefile is a geospatial vector data format mainly used in Geographic Information Systems (GIS) \citep{shapefile}. Such files are useful to obtain information about country geometry described using polygons and multipolygons.  A shapefile of each country with an incomplete FM transmitter database was obtained from GADM\footnote{\href{https://gadm.org/}{GADM.org}}. Since HEALPix works with raster data, the shapefile was converted into a high resolution raster using the `rasterize' function in QGIS \citep{QGIS_software} and converted into a HEALPix map, thereby obtaining all the HEALPix pixels within that country of selected \NSide{} resolution. The 2-D spatial distribution of the incomplete dataset was then projected onto the HEALPix maps using a standard multi-dimensional interpolation scheme. 

Synthetic transmitter locations were then generated by randomly sampling pixel positions within each country according to its normalized spatial probability distribution. The number of sampled points corresponded to the transmitter deficit for that country.



The statistical model was tested for the incomplete USA dataset against the already obtained complete USA dataset as mentioned in Section \ref{subsec: Incomplete datasets} (See Figure \ref{fig:dist_comp}). The synthetic transmitter HEALPix map and the original transmitter HEALPix map were binarized- meaning any pixel having at least one FM transmitter would have a value of 1, while any pixel having no FM transmitters would have a value of 0. Performance of the model was measured using accuracy, precision and recall (see Equation \ref{eq: acc})  \citep{recovering_wedge}. \par
\begin{equation}\label{eq: acc}
\begin{split}
\text{Accuracy} &= \frac{TP + TN}{TP + TN + FP + FN}, \\
\text{Precision} &= \frac{TP}{TP + FP}, \\
\text{Recall} &= \frac{TP}{TP + FN},
\end{split}
\end{equation}
where TP, TN, FP and FN stand for True Positive, True Negative, False Positive, and False Negative respectively. The performance of the model with different sampling resolutions were tested by varying the \NSide{} of the maps that were being sampled.\par  
\par
As previously mentioned, a low sampling resolution would result in underfitting, and a high sampling resolution would result in overfitting, forcing synthetic transmitters to be placed at the  same location as the sample. Hence, an optimal sampling resolution would be a value of \NSide{} that lies between these two extrema and has the best performance metrics. Looking at the performance of the model as seen in table \ref{tab:model_performance}, a sampling resolution of \NSide{} 32 (and subsequently lower \NSide{} values) seems to have the best overall performance. However, this is attributed to the fact that a low-resolution binarized FM transmitter map will mostly have pixels with the value 1. This leads to a drastic increase in the number of TPs thus falsely returning a very high performance score (see Equation \ref{eq: acc}), even if the synthetic transmitter locations did not reflect real locations with accuracy. Thus, \NSide{} 64 was selected to be the optimal resolution for sampling as it had the best verifiable performance metrics while having an intermediate spatial resolution.

\begin{table}[h]
\caption{Performance of the model for different \NSide{} HEALPix maps.}
\label{tab:model_performance}
\centering
\begin{tabular}{cccc}
\hline
\NSide{} & Accuracy & Precision & Recall \\
\hline
32  & 0.83 & 0.93 & 0.86 \\
64  & \textbf{0.80} & \textbf{0.66} & \textbf{0.66} \\
128 & 0.80 & 0.48 & 0.37 \\
256 & 0.87 & 0.19 & 0.28 \\
512 & 0.94 & 0.06 & 0.15 \\
\hline
\end{tabular}
\end{table}

\subsubsection{Modeling FM Transmitter frequency parameter}\label{subsubsec: freq param}
The frequency of the synthetic transmitter will be higher than the last recorded frequency in the incomplete transmitter dataset, as FMLIST presents data in ascending order of frequency. Therefore, there were no existing spatially dependent distribution functions to sample for the missing frequencies. Hence, frequency parameter was assumed to be spatially independent thus making it a univariate distribution. We use the PRATUSH window of 55-110 MHz to sample synthetic transmitter frequency values.\par
New frequency values were generated using `reverse-interpolation' sampling.
Reverse-interpolation sampling is done by creating a normalized cumulative distribution function (CDF) for the univariate distribution. New samples can be drawn from the normalized CDF by reverse-interpolating a uniform distribution ranging from 0 to 1 with the CDF  to obtain the corresponding values of the variable. We use the Freedman-Diaconis rule to bin frequency values \citep{freedman1981histogram} given by Equation \ref{eq : FDrule}

\begin{equation}\label{eq : FDrule}
h = \frac{2 \cdot IQR(x)}{\sqrt[3]{n}},
\end{equation}
where \( h \) is the bin width, \( IQR(x) \) is the inter-quartile range of the data, and \( n \) is the number of data points. The frequencies were randomly sampled and assigned to each synthetic transmitter independent of their location parameter. 

\subsubsection{Modeling FM transmitter EIRP parameter}
The spatial dependence of FM transmitter EIRP can be measured using the `Local Moran's I test' for spatial autocorrelation \citep{Moran} given by Equation \ref{eq:Morans} 
\begin{equation}\label{eq:Morans}
I_i = \frac{(x_i - \bar{x})}{S^2} \sum_{j=1}^n w_{ij} (x_j - \bar{x}),
\end{equation}
where $I_i$ is the local Moran's I value at observation $i$, $w_{ij}$ is the spatial weighting factor between observation $i$ and nearest $j$ neighbors, $x_i$ is the value of the transmitter EIRP, $x_j$ is the transmitter EIRP value at location $j$, $\bar{x}$ is the average transmitter EIRP in that country, and 
$S^2$ is the global transmitter EIRP variance given by \begin{equation}
    S^2 = \frac{\sum_{i=1}^{n} (x_i - \bar{x})^2}{N},
\end{equation}\par
\begin{figure}
    \centering
    \includegraphics[width=0.9\linewidth]{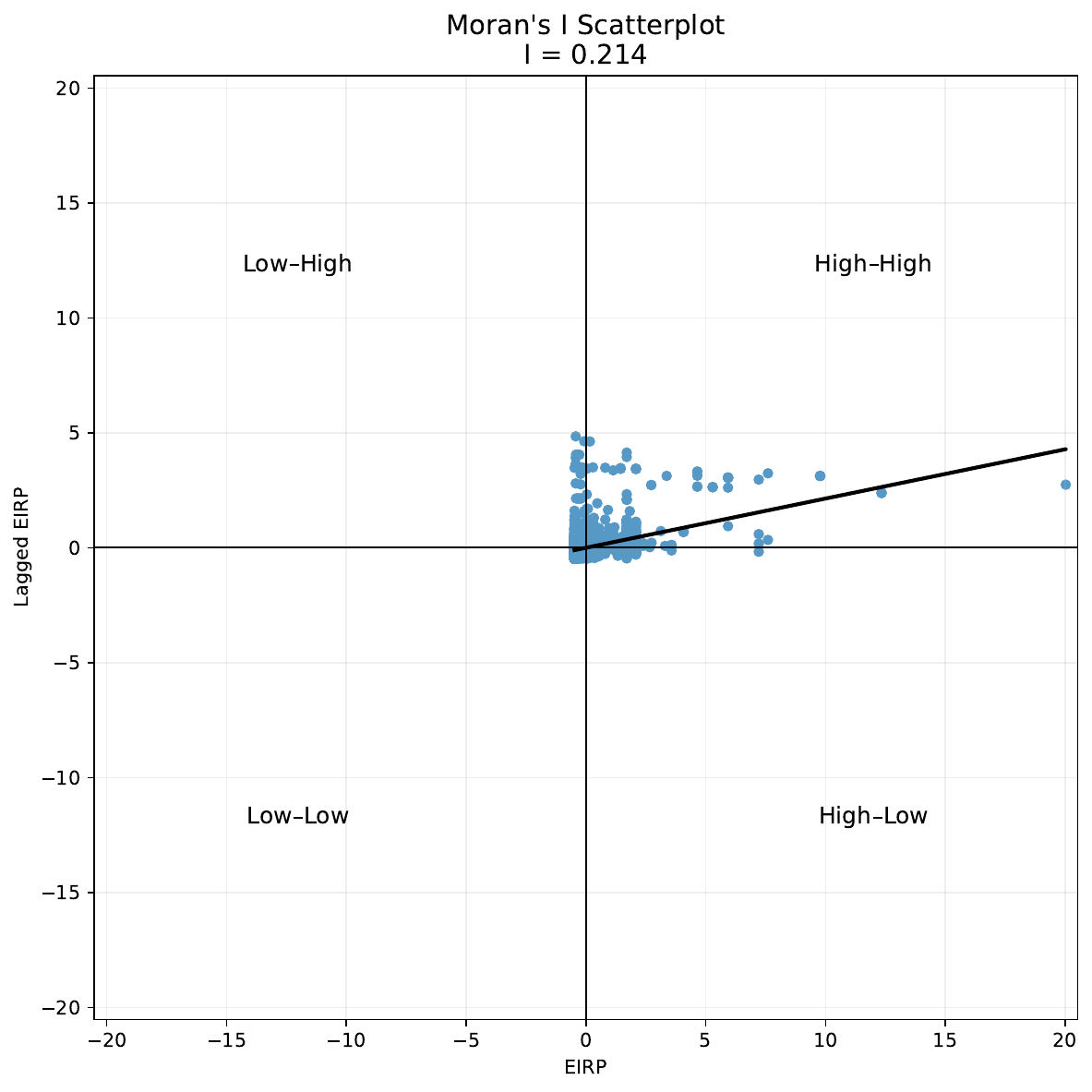}
    \caption{Local Moran's I scatter plot of FM transmission EIRP for Russia}
    \label{fig:Moran scatter}
\end{figure}

Since the local Moran's I is an inferential statistic, we compute the statistical significance of the obtained local Moran's I value with hypothesis testing. We form a null hypothesis stating that the transmitter EIRP values are NOT spatially correlated at $i^{th}$ observation with $j$ neighbors. We then infer the spatial dependence of transmitter EIRP distribution with the obtained Z-score given by Equation \ref{eq:Z moran}
\begin{equation}\label{eq:Z moran}
    Z_i = \frac{I_i - \mathbb{E}[I_i]}{\sqrt{\text{Var}(I_i)}},
\end{equation}
where $Z_i$ is the Z-score at observation $i$, $I_i$ is the local Moran's I value obtained at observation $i$, $\mathbb{E}[{I_i}]$ is the expected local Moran's I value at $i$ and is given by Equation \ref{eq:E moran}
\begin{equation}\label{eq:E moran}
    \mathbb{E}[I_i] = -\frac{\sum_{j=1}^n w_{ij}}{n - 1},
\end{equation}
where $n$ is the number of transmitters in the incomplete dataset and $\text{Var}(I_i)$ is the variance of the local Moran's I under the null hypothesis and is given by 
\begin{equation}
    \text{Var}(I_i) = \mathbb{E}[I_i^2] - \mathbb{E}[I_i]^2.
\end{equation}
The local Moran's I was computed for each country with incomplete dataset for transmitter EIRP's spatial autocorrelation locally along with the statistical significance of the value in QGIS. The nature of spatial autocorrelation of transmitter EIRP can be analyzed by categorizing the local Moran's I value into four different quadrants visualized in a Moran's I scatter-plot as shown in Figure \ref{fig:Moran scatter}. It presents the spatial autocorrelation by plotting the standardized average transmitter EIRP value of `nearest j-neighbors' referred to as `lagged EIRP' against the standardized transmitter EIRP value. A straight line fit of global Moran's I index is also shown. The first quadrant denotes high-high autocorrelation (EIRP at $i$ location is high; average EIRP of nearest $j$ neighbors are also high and so on) , the second quadrant denotes high-low autocorrelation, the third quadrant denotes low-low autocorrelation and the fourth quadrant denotes low-high autocorrelation. The first and third quadrants represent clustering, and the second and fourth quadrants represent dispersion. Most of the transmitter EIRP values are clustered toward the first quadrant implying that a majority of the transmitter EIRP values have `high-high' spatial autocorrelation. This was found to be the case for all countries with incomplete datasets.\par

\begin{figure}[ht]
  \centering
  \includegraphics[width=0.89\linewidth]{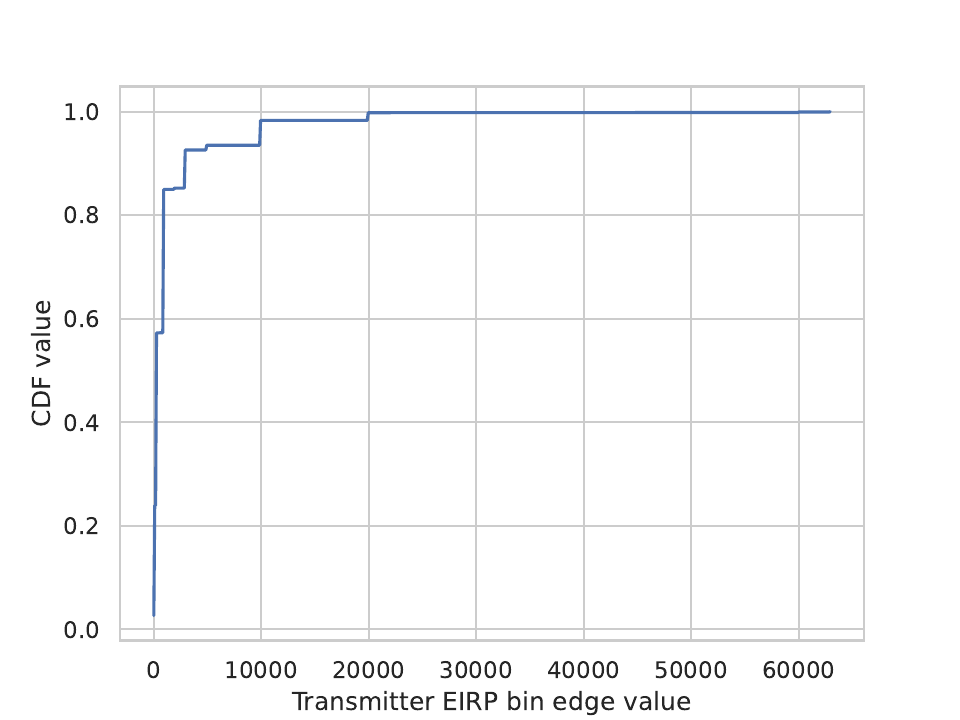}
  \caption{Cumulative function distribution of EIRP of FM Tx for an incomplete dataset}
  \label{fig:CDF EIRP}
\end{figure}

For each new transmitter that was populated, first the transmitter location and frequency were assigned following the procedures described in Sections \ref{subsubsec: location} and \ref{subsubsec: freq param}. Then a mapping was created between the HEALPix pixel of the transmitter and the quadrant where its corresponding local Moran's I value was situated. New EIRP values were assigned to the synthetic transmitters by reverse interpolation sampling with EIRP values sampled from the corresponding quartiles of the transmitter EIRP CDF. For example, if a synthetic transmitter was located in a pixel which had its local Moran's I value in the first quadrant (which corresponds to high-high autocorrelation), the synthetic EIRP value was sampled from the fourth quartile of the transmitter EIRP CDF (See Figure \ref{fig:CDF EIRP}). If the synthetic transmitter lied in a region where there was no spatial autocorrelation (Statistically insignificant local Moran's I value; meaning it would lie at the center of the local Moran's I scatterplot), the synthetic transmitter EIRP was assigned by drawing random samples from the entire CDF of transmitter EIRP. 

\begin{figure}[ht]
  \centering
  \includegraphics[width=0.89\linewidth]{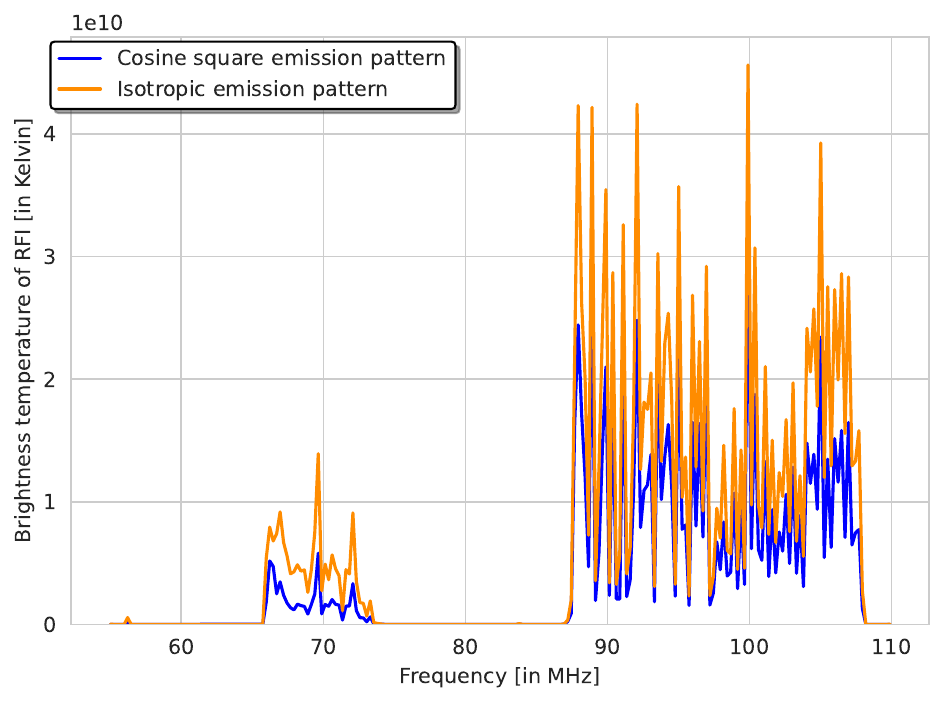}
  \caption{RFI spectrum over India with the new STARFIRE-2 database with two radiation patterns used by the FM transmitters.Isotropic radiation pattern has higher RFI brightness temperature thus making it a conservative assumption for PRATUSH.}
  \label{fig:India_newspec}
\end{figure}

\begin{figure*}
    \centering
     \begin{subfigure}[b]{0.45\textwidth}
        \centering
        \includegraphics[width=0.89\linewidth]{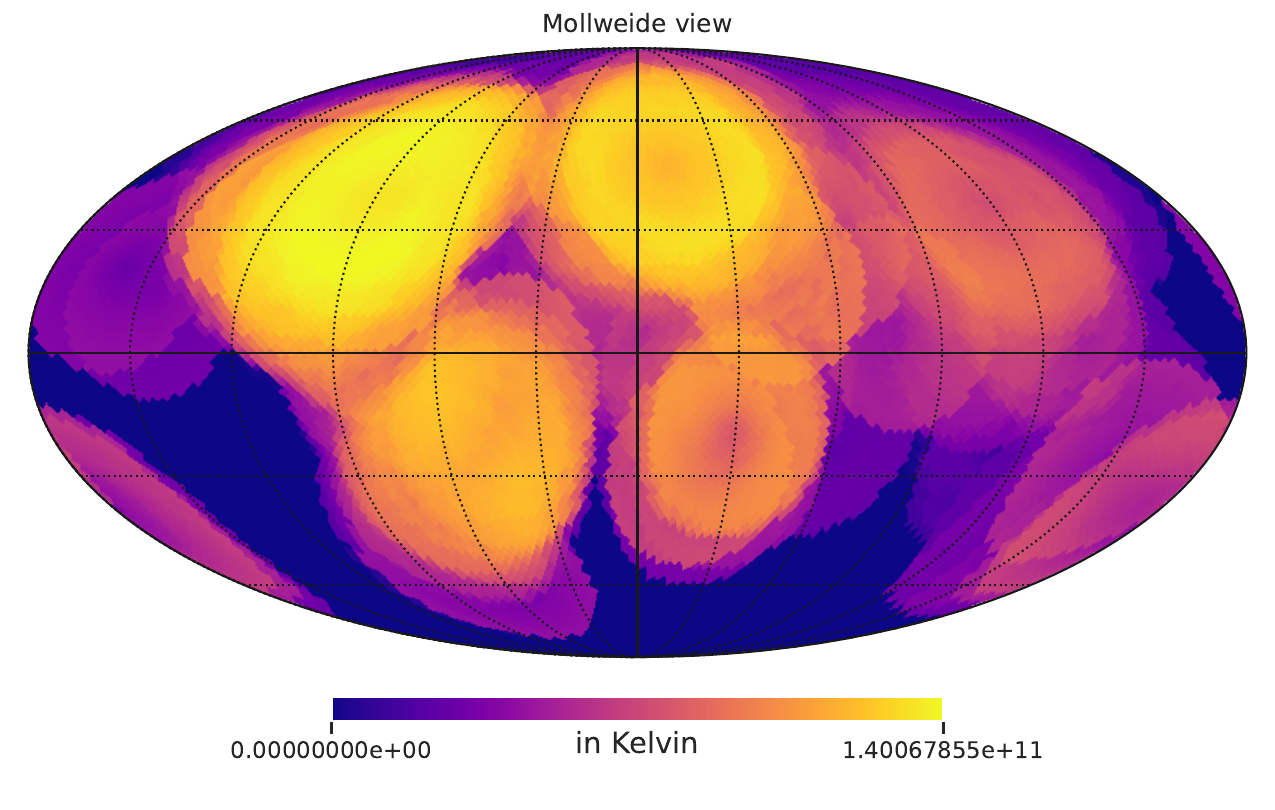}
        \caption{Beam averaged pixel-wise RFI power at 400 km altitude and frequency 107.22 MHz with new STARFIRE-2 database.}
        \label{fig:newHM}
    \end{subfigure}
    \begin{subfigure}[b]{0.45\textwidth}
        \centering
        \includegraphics[width=0.89\linewidth]{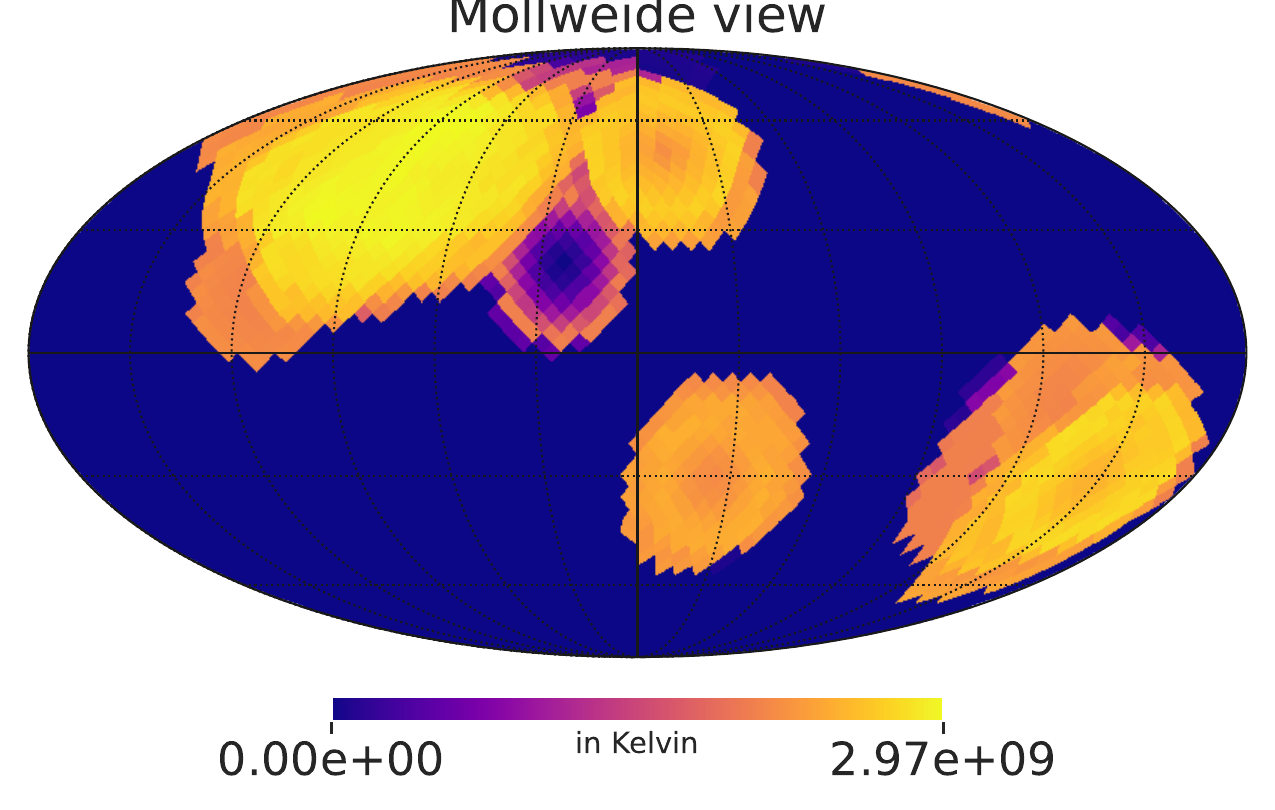}
        \caption{Beam averaged pixel-wise RFI power at 400 km altitude and frequency 107.22 MHz with original STARFIRE database.}
        \label{fig:oldHM}
    \end{subfigure}
    \hfill
   
    \caption{Comparison of beam-averaged pixel-wise RFI power at 400 km altitude for different STARFIRE database versions.}
    \label{fig:heatmap_comparison}
\end{figure*}

\begin{figure}
  \centering
  \includegraphics[width=0.89\linewidth]{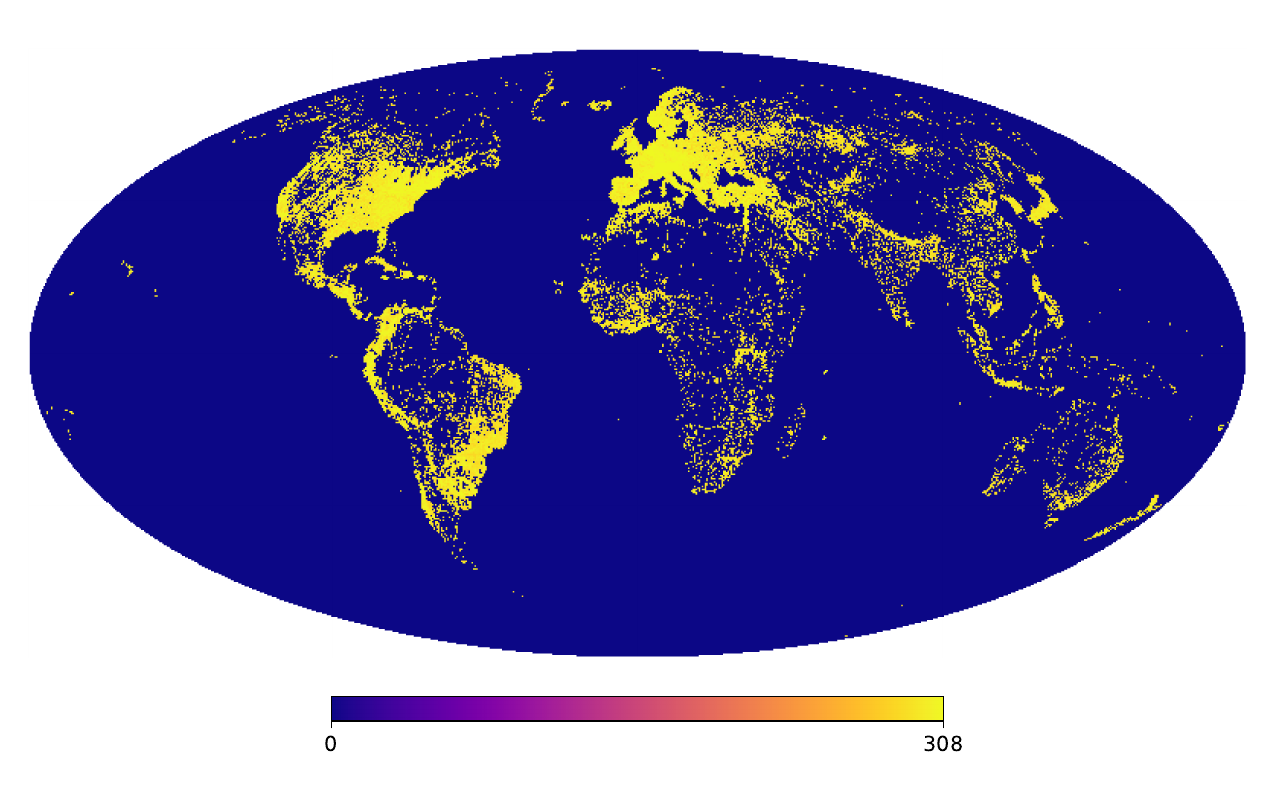}
  \caption{ \NSide{} 128 map of number of transmitters per pixel with a complete database}
  \label{fig:Tx no w/ complete}
\end{figure}
\subsection{The complete FM transmitter database}\label{subsec: Complete db}
We henceforth call the STARFIRE algorithm with the newly completed database as STARFIRE-2. Changes in the RFI spectral cube after running the newly completed FM transmitter database are shown in Figure \ref{fig:heatmap_comparison}. The RFI spectra seen by PRATUSH over Hanle (India) with different transmitter radiation patterns with the newly completed STARFIRE database is shown in Figure \ref{fig:India_newspec}. The older version of STARFIRE did not predict any RFI over this region due to incompleteness in its database. The complete database as a HEALPix map of \NSide{} 128 is seen in Figure \ref{fig:Tx no w/ complete}.
 The algorithm was also run using a $cos^2(\theta)$ radiation pattern for the FM transmitters, with $\theta$ being the elevation angle of the satellite with respect to the FM transmitter. This could be a more realistic assumption for radiation pattern used by FM transmitters to maximize the range of signal propagation on Earth. As seen in Figure \ref{fig:India_newspec}, isotropic radiation pattern for FM transmitters have higher RFI levels in comparison to horizon-directed beams.

\section{Orbit simulations}\label{sec: orbit sims}
\begin{figure}
  \centering
  \includegraphics[width=0.89\linewidth]{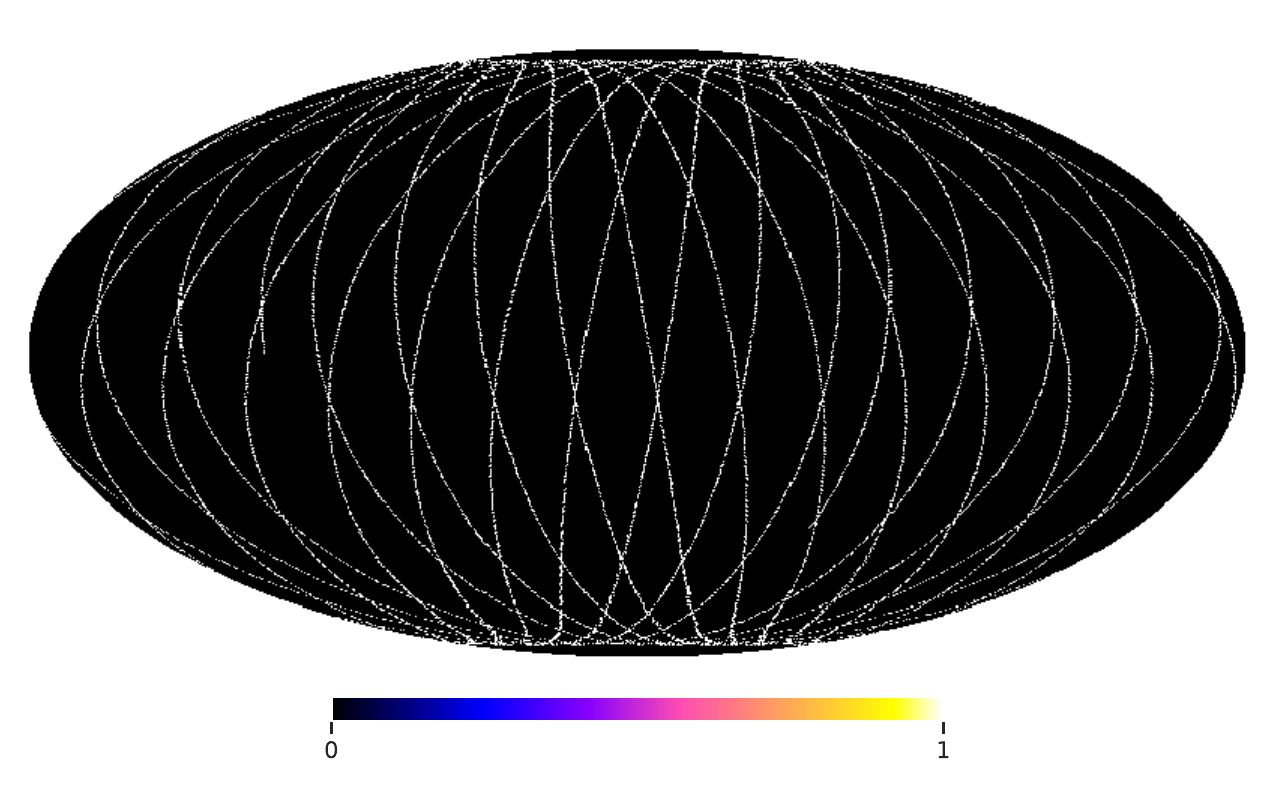}
  \caption{Ground-track of one day of orbit propagation for a polar orbit at 705 km altitude}
  \label{fig:GT plot}
\end{figure}

The database was completed using the methods mentioned in the previous section. This complete database was run through the STARFIRE-2 algorithm with the following specifications:
\begin{itemize}
    \item [--] Frequency range of receiver: 55-110 MHz
    \item [--] Frequency resolution of the receiver: 244 KHz
\end{itemize}
To estimate the RFI in Earth orbit, we propagate the satellite position over Earth using orbital parameters of interest and compute the expected RFI as the satellite moves over pixels corresponding to the "ground-track"- the path that the satellite traces on the surface of the Earth directly beneath while in orbit (See Figure \ref{fig:GT plot}). For each pixel in the ground-track, we estimate the corresponding RFI spectrum from the appropriate slicing of the STARFIRE-2 spectral cube.\par  
Orbit simulations were done by using `Two-Line Element set' (TLE) of different satellite orbits. TLE is a data format used to list orbital elements of a satellite at a given epoch. Orbits are propagated with the TLE information using the SGP4 (Simplified General Perturbation) model \citep{whatTLE}. The orbit propagation algorithm uses the `Skyfield' Python library  \citep{2019ascl.soft07024R} and returns the ground-track coordinates as a function of time.\par
The quality of an orbit can be assessed using the ``Figure of Merit"(FOM), a metric introduced in \citet{STARFIRE}. The FOM quantifies the number of frequency channels free from any Radio Frequency Interference (RFI), with all RFI-contaminated spectral channels flagged irrespective of the interference power in those channels. Each FOM bin contains 4 frequency channels in the case of PRATUSH-I. Based on the FOM, regions where the satellite experiences minimal or no RFI are designated as low RFI zones. Specifically, any pixel with an FOM value of 54 or higher (with 55 being the maximum possible FOM) is classified as a low RFI zone. To evaluate different orbital configurations, the total time the satellite spends within these zones is calculated. An optimal orbit is therefore one that maximizes the time spent in high-FOM regions, or equivalently, in zones characterized by low levels of RFI contamination.

We choose one year as the orbit propagation period. However, orbit propagation errors using SGP4 models grow quadratically with propagation time for Low Earth Orbits (LEO) \citep{TLE_error}. A feasible method to simulate an orbit would be to obtain archival TLE information from missions that have been active for at-least one year, since historical TLE files for LEO satellites have a propagation error of $\pm$ 0.25 km for one day of orbit propagation \citep{historical_TLE}. This is well within the minimum resolution of the HEALPix maps ($\sim$ 205 km at the equator) used for RFI analysis. Therefore, in order to maintain accurate ground-track throughout the span of orbit propagation, TLE data for  select satellites (see Section \ref{subsec:orbits} for details) were obtained from NORAD archive\footnote{\href{https://celestrak.org/NORAD/archives/}{Celestrak.org}} such that the TLE files would be renewed every day throughout the span of one year of orbit propagation. 

The time resolution for recording ground-track coordinates of the satellite in orbit was set to 8.640 seconds. Finally the total time spent by the satellite at each FOM bin was calculated for only regions and times when the satellite would not be sunlit since PRATUSH-I would conduct science observations when the sun is below the horizon (night time) to avoid intercepting incoming solar radio flux.\par

\subsection{South Atlantic Anomaly (SAA)}
There exists a region over Earth where its magnetic field strength is the weakest. This is due to the offset of the Earth's magnetic dipole from Earth's center \citep{SAAcause}. Due to the offset of the two axes (rotational and magnetic), the Van Allen radiation belt containing high energy particles can penetrate deeper into the atmosphere (altitudes of 200 km from the Earth surface) at the SAA region. This effect causes satellites in Low Earth Orbit (LEO) to receive large doses of radiation in orbit \citep{SAA_dosage} and potentially lead to `Single Event Upsets' which causes electronics to malfunction after interacting with a high energy particle \citep{SEU}. The SAA region spans from 90\textsuperscript{$\circ$} W to 40\textsuperscript{$\circ$} E longitude and 0\textsuperscript{$\circ$} to 50\textsuperscript{$\circ$} S latitude \citep{SAA_extent}.\par
We regard this as an unfavorable region for conducting observations.
 We account for this by setting the FOM value to 0 over this region, thereby down weighting any pixels that lie in the SAA which might otherwise have a high FOM. 

\subsection{Optimal orbit selection}\label{subsec:orbits}
Orbits of three satellites were simulated as representative to evaluate the time spent by the satellite in low RFI zones - two of which were in Low Earth Orbit (LEO) and a third one in geosynchronous orbit. A polar LEO was simulated using archival TLE data of `Landsat-7' \citep{Landsat7} and an equatorial LEO was simulated using archival TLE data of `AstroSat' \citep{astrosat}. `NSS-9' data was used to simulate  geosynchronous orbit.\par
\begin{figure}
    \centering
    \includegraphics[width=0.89\linewidth]{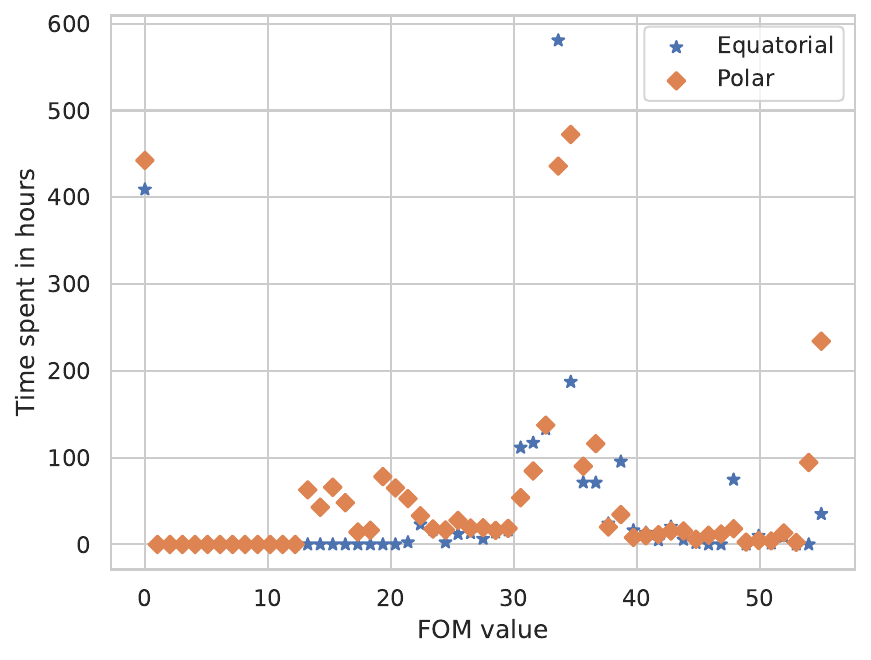}
    \caption{Time spent by the satellite at each FOM bin for an orbit propagation of 1 year}
    \label{fig:orbit time compare}
\end{figure}
As seen in Figure \ref{fig:orbit time compare}, the polar orbit outperforms the equatorial orbit for the most amount of time spent over low RFI zones for an orbit-propagation period of one year (234.582 hours a year). This is attributed to the fact that the satellite in polar orbit spends more time in radio-quiet zones including the Pacific ocean, Antarctica, southern Atlantic and Indian oceans. In contrast, a satellite in equatorial orbit would mostly spend its time over RFI contaminated zones (it spends 35 hours in low RFI zones per year). Geosynchronous orbits have zero hours of low RFI zone time for a year of orbit propagation since the FOV of the satellite covers a hemisphere of the Earth thereby containing multiple RFI sources in its beam at any location over the Earth. As mentioned in section \ref{subsec: Complete db}, the orbit simulations were also run setting the radiation pattern of the FM transmitters to a $cos^2(\theta)$ pattern, with $\theta$ being the elevation angle. The time PRATUSH spends at radio quiet zones changed to 253.2 hours per year in polar orbit, and remained unchanged for the equatorial orbit. With these results, we choose to set the radiation pattern of the FM transmitters to be isotropic to have conservative estimates for our sensitivity estimates in the next section of this paper.\par

\begin{figure}
    \centering
    \includegraphics[width=\linewidth]{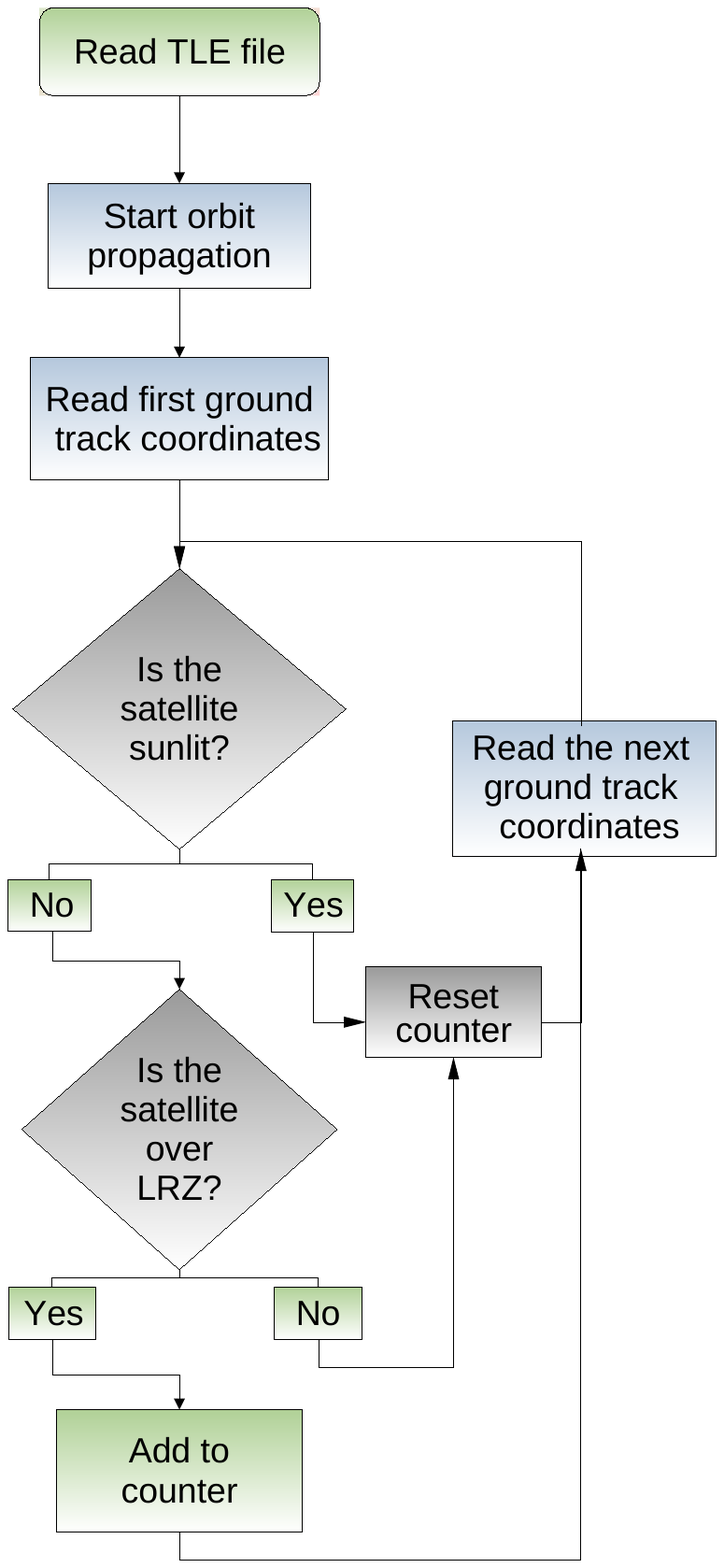}
    \caption{Uninterrupted low-RFI counter algorithm, where `LRZ' stands for Low RFI Zone.}
    \label{fig:Flowchart}
\end{figure}

\begin{figure}[h]
    \centering
    \begin{subfigure}{0.89\linewidth}
        \centering
        \includegraphics[width=\linewidth]{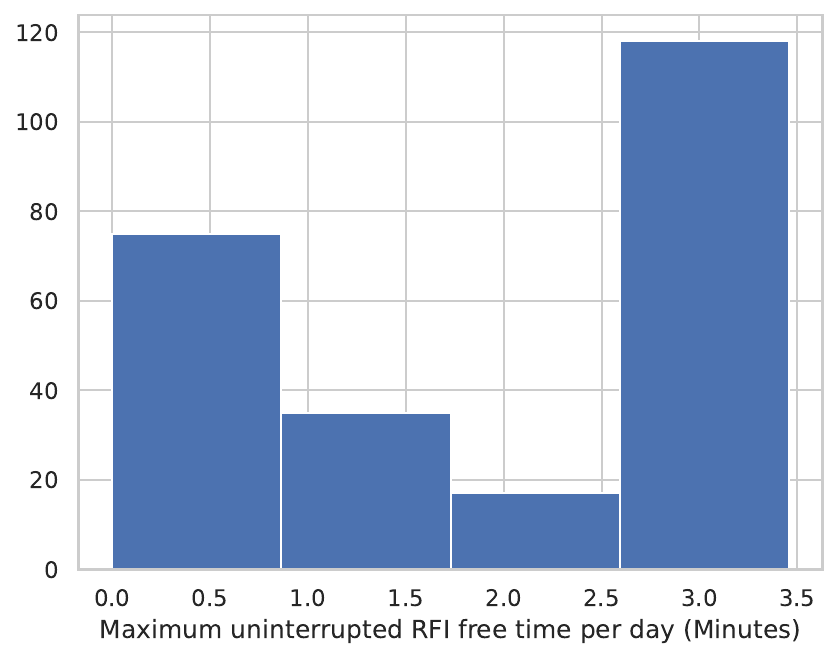}
        \caption{Maximum Uninterrupted low RFI zone time per day in equatorial orbit}
        \label{fig:equi unint}
    \end{subfigure}

    \vspace{0.5em}

    \begin{subfigure}{0.89\linewidth}
        \centering
        \includegraphics[width=\linewidth]{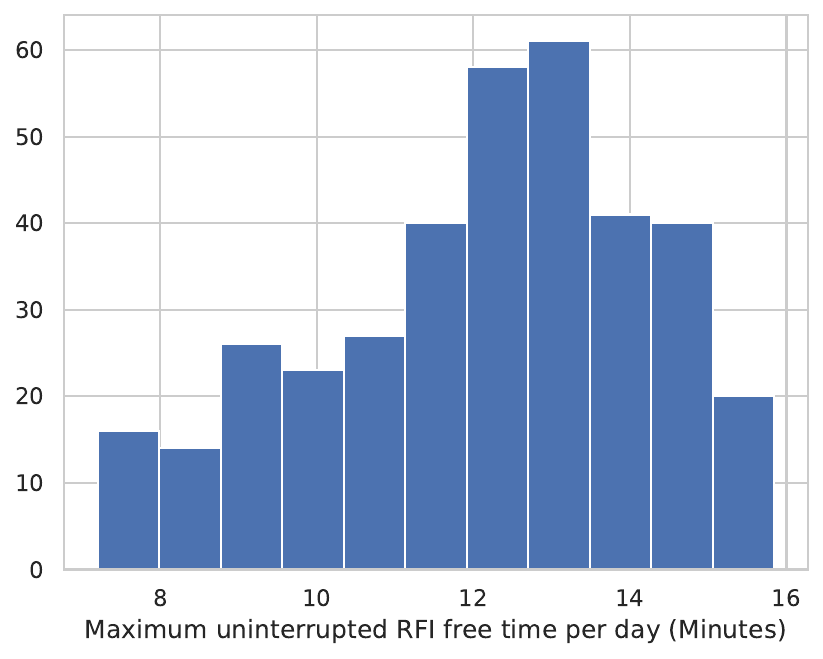}
        \caption{Maximum Uninterrupted low RFI zone time per day in polar orbit}
        \label{fig:pol unint}
    \end{subfigure}

    \caption{Maximum uninterrupted low-RFI zone time per day for equatorial and polar orbits.}
\end{figure}

We also investigate the transit periods over low RFI zones the satellite has for different orbits per day. The orbit simulation algorithm records uninterrupted low-RFI time. The methodology  is visualized in Figure \ref{fig:Flowchart}. 
Figures \ref{fig:equi unint} and \ref{fig:pol unint} show the distribution of maximum uninterrupted low-RFI time per day 
in minutes for a satellite in polar and equatorial orbit respectively.
As expected, a satellite in polar orbit performs better in this parameter as well in contrast to an equatorial orbit.

\section{Mock-sky observations}\label{sec: sky obs}
The sensitivity of PRATUSH-I to various CD/EoR models can be estimated by including mock observations of the radio sky included in the STARFIRE-2 algorithm with the complete database. This was done by obtaining the beam weighted sky for PRATUSH-I antenna given by Equation \ref{eq:BWS} :

\begin{equation}
\label{eq:BWS}
\begin{aligned}
T_{\mathrm{A}}(\nu)
&= \bigl(1 - |S_{11}(\nu)|^{2}\bigr)
\Bigg[
\frac{
\displaystyle
\int_{0}^{2\pi}\!\!\int_{-\pi/2}^{0}
T_{\mathrm{RFI}}(\theta,\phi,\nu)\,
B(\theta,\phi,\nu)\,{\rm d}\Omega
}{
\displaystyle
\int_{0}^{2\pi}\!\!\int_{-\pi/2}^{0}
B(\theta,\phi,\nu)\,{\rm d}\Omega
}
\\[6pt]
&\qquad+
\frac{
\displaystyle
\int_{0}^{2\pi}\!\!\int_{0}^{\pi/2}
T_{\mathrm{Sky}}(\theta,\phi,\nu)\,
B(\theta,\phi,\nu)\,{\rm d}\Omega
}{
\displaystyle
\int_{0}^{2\pi}\!\!\int_{0}^{\pi/2}
B(\theta,\phi,\nu)\,{\rm d}\Omega
}
+ T_{\mathrm{Earth}}(\nu)
\Bigg],
\end{aligned}
\end{equation}

where $T_{\mathrm{A}}$ is the antenna temperature, $S_{11}(\nu)$ is the reflection coefficient of the PRATUSH antenna, $T_{\mathrm{RFI}}$ is the RFI temperature that PRATUSH intercepts from the lower half of the azimuthal plane, $T_{\mathrm{Sky}}$ is the foregrounds brightness temperature from GMOSS maps \citep{GMOSS}, $B$ is the beam pattern of the PRATUSH antenna, and $T_{\mathrm{Earth}}$ is the effective brightness temperature of the Earth approximated as a black body of temperature 300K \citep{EarthTemp}.  

In addition to calculating the elevation angle an FM transmitter subtends with respect to the satellite, one requires the azimuthal angle subtended by the transmitter with respect to the subsatellite point.
The azimuth angle of a ground point with respect to the sub-satellite point is defined as
\begin{equation}
\text{Azimuth} = \arctan2 \!\left( \mathbf{v} \cdot \hat{e}, \; 
\mathbf{v} \cdot \hat{n} \right),
\end{equation}
where $\mathbf{v} = \mathbf{r}_g - \mathbf{r}_{\text{ssp}}$ is the 
line-of-sight vector from the sub-satellite point to the ground point, and 
$\hat{e}, \hat{n}$ are the local east and north unit vectors in the ENU 
coordinate system. \citep{seeber2003satellite}

\par

\subsection{Sensitivity to CD/EoR models}
\begin{figure*}
    \centering
    \includegraphics[width=\linewidth]{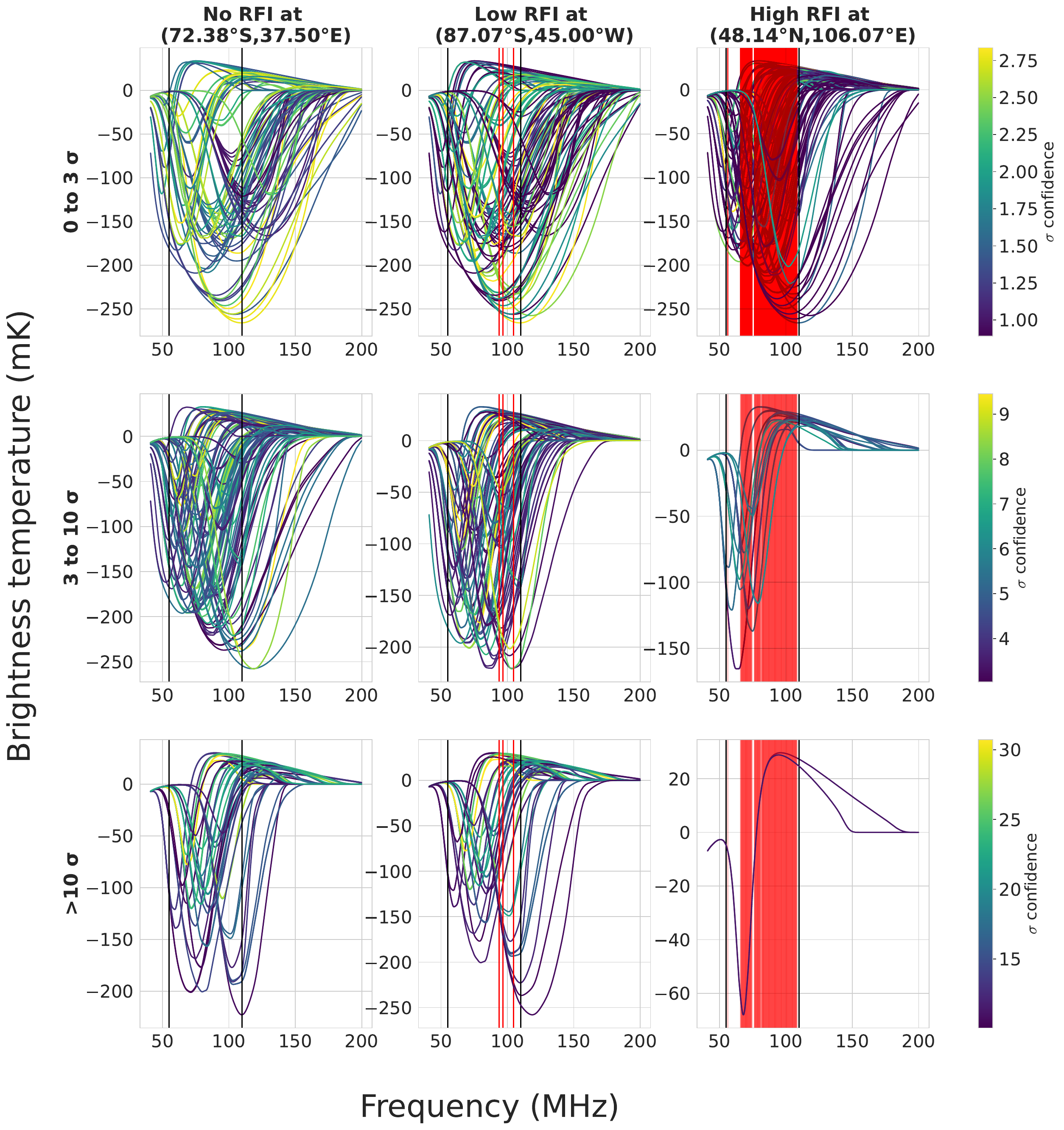}
    \caption{Confidence in detecting different CD/EoR signal models at different locations. Each column of the panel represents an RFI scenario encountered in the optimized orbit, while each row represents a confidence interval. The solid black lines indicate the PRATUSH-I frequency window, and the solid red lines indicate frequency channels lost to RFI. }
    \label{fig: mega plot}
\end{figure*}

We examine the sensitivity of the radiometer for different CD models at various points in the orbit corresponding to varying levels of FM-seeded RFI contamination. We compare three scenarios, a region over Earth with zero channels lost to RFI ((72°23'15"S  37°30'00"E) over Antarctica), a region with four channels lost / flagged ((87°4'33"S 
 45°0'0"W) over Antarctica), and a region with very high RFI levels (48°8'28"N  106°4'17"E) leading to high channel loss (153 channels lost out of 226 channels).  A threshold based flagging routine was employed to flag channels with very high brightness temperature. From GMOSS, we can expect a galactic contribution of $\sim$ 6000 K at 55 MHz with the beam maxima pointing towards the galactic center. Hence we round up the threshold for flagging any channel that contains more than 10,000 K as a conservative estimate. We do not explore more complex RFI flagging routines beyond threshold flagging as it is beyond the scope of this paper. Sensitivity to various CD/EoR models at different regions over the Earth is shown in Figure \ref{fig: mega plot}.
Foregrounds are expected to spectrally smooth in contrast to the global signal and can be modeled by a maximally smooth function \citep{MS_func}. Foregrounds are separated by subtracting the best fit maximally smooth function and minimizing the residual root mean square (RMS) using the `Basinhopping' algorithm \citep{Bhopping}. Various CD/EoR global-signal models \citep{CD_mods} were injected into the beam weighted sky and were separated from the foregrounds using the maximally smooth function. Confidence in detecting each model was calculated by comparing the increase in the residual RMS for the simulated sky spectrum with the 21-cm signal injected and that for a mock-sky with no signal injected, which is a null hypothesis case. The noise-floor for conducting the mock sky observations was determined by the radiometer equation 
\begin{equation}\label{eq:Radiometer}
    \Delta T = \frac{T_\mathrm{sys}}{\sqrt{B \, \tau}},
\end{equation}
where $\Delta T$ represents the radiometer noise temperature, $T_\mathrm{sys}$ represents the system temperature, $B$ is the channel width of the instrument, and $\tau$ is the integration time, which was obtained by calculating the total RFI free time for a year of orbit propagation in the polar orbit. \citep{ERA}

\begin{table}
\caption{Percentage of detecting different CD/EoR models at different confidence levels.}
\label{tab:conf_lvls}
\centering
\begin{tabular}{lccc}
\hline
Confidence Level & No RFI & Low RFI & High RFI \\
\hline
0--3 $\sigma$   & 34.6\% & 34.6\% & 90.9\% \\
3--10 $\sigma$  & 45.6\% & 45.6\% & 8.3\%  \\
$>$10 $\sigma$  & 19.7\% & 19.7\% & 0.76\% \\
\hline
\end{tabular}
\end{table}

As seen in Table \ref{tab:conf_lvls}, having a small number of channels lost to RFI does not have an impact on the confidence in the detection of  various CD/EoR models. In contrast, high channel loss leads to a majority of the CD/EoR models not being detected with high confidence. With these results, we can conclude once again that any orbit that spends the most amount of time in low RFI zones would be deemed optimal- the polar LEO in this case. 


\section{Discussion} \label{sec: Discussion}
\subsection{PRATUSH at L2 point of the Sun-Earth system}
One can investigate the feasibility of conducting such an
experiment much farther from the Earth to see if RFI
power falls below the amplitude of the 21-cm signal. One possibility is the Sun-Earth L2 point where many astronomy telescopes have flown in an Lissajous orbit \citep{JWST,hechler2003herschel}. For simplicity of the feasibility study, we place PRATUSH at the L2 point, rather than in orbit around L2, in order to estimate the order of magnitude of the RFI power levels contributed by the Earth-facing hemisphere. Incoming radio flux from the Sun
due to various time varying solar activities are not taken
into account but will surely be detected by the experiment even in an ideal case where the satellite would
be eclipsed from the Sun, since the Sun’s apparent size at
L2 is slightly larger than the Earth (by around $0.022^\circ$).
\begin{figure}
    \centering
    \includegraphics[width=\linewidth]{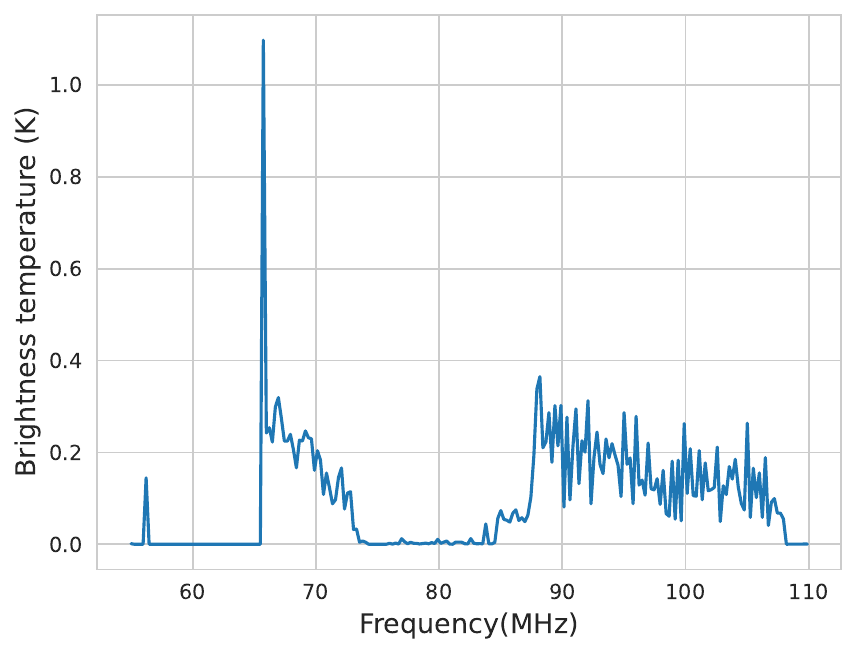}
    \caption{RFI power levels at L2 of the Sun-Earth system}
    \label{fig:L2}
\end{figure}
As seen in Figure \ref{fig:L2}, the RFI brightness temperature
lies in the order of a few hundred mK, going up to a Kelvin, thus adding significant
spectral features to the sky spectrum. This would result
in poor sensitivity to all CD/EoR models.

\subsection{Sources of uncertainty}
\begin{itemize}
    \item All simulations demonstrated in this paper are with the caveat that the database procured from FMLIST is a reasonably accurate representation of FM transmission. There are a few discrepancies including missing information.  18.75\% of transmitters obtained from the FMLIST database have missing EIRP information, which were replaced with power values drawn from a global transmitter power distribution. However the interpretation of the STARFIRE-2 algorithm using the FOM metric is insensitive to the exact value of EIRP introduced since flagging is based on presence or absence of RFI in a channel and not the RFI power. 0.35\% transmitters have missing location information that are filled using the statistical methods described herein. While non-zero, this is not a vary large number and we acknowledge this to be a source of uncertainty.
    \item The exact location, frequency, power of the missing database entries populated using statistical methods are representative and reflect the underlying statistical distribution of the population.
    \item Shapefiles being used for generating synthetic transmitter information can have missing spatial information (Archipelagos, small territories with lesser angular extent) thus affecting the performance of the probabilistic model of transmitter distribution.
    \item The radiation pattern of all FM transmitters are assumed to isotropic to reduce the complexities involved in obtaining the beam pattern of each FM transmitter. This is a conservative assumption compared to omni-directional antennas that are typically employed by FM transmitters. In practice the transmitting antennas may be more directive due to which there may be more number of cleaner patches in the output heatmaps of STARFIRE-2 and thus more locations in orbit that are classified as low RFI zones.

    \item Other sources of RFI in the band, including and not limited to dynamic sources, such as other satellites in orbit, maritime radars, harmonics and wideband noise from power systems from other sources, are not included in the current database. We also do not include unintended and hence undocumented sources of emission, such as from power systems onboard satellites that have been observed by low frequency telescopes on Earth \citep{zhang2025broadband, bassa2024bright, di2023unintended}. These are best determined empirically due to the unregulated nature of emission, which motivates a dedicated measurement of the radio environment in orbit around Earth.
    \item The results presented in the paper will be dramatically different for an antenna that has zero/high suppression beam response in the lower half of the azimuthal plane. For instance, it is possible to achieve a narrow beam that is pointed away from Earth, say via beamforming, causing significant suppression in RFI levels. Not only would such an antenna be very large to achieve the directive beam at the frequencies in question, achieving a stable achromatic beam over the full frequency range is extremely difficult. Thus, in the case of PRATUSH-I, antenna design considerations include frequency independent behavior in the band of are prioritized, making a narrow-beam response not plausible.
    \item The statistical methods employed in this paper inevitably introduce a positive bias. Consequently, the model might fail to populate locations with missing transmitters if none were present in the prior distributions, even though they may exist in reality.
\end{itemize}

\section{Conclusions and future work} \label{sec: Conclusions}
It is worth exploring an Earth orbiting radiometer as a precursor to a lunar farside experiment for detecting the global redshifted 21-cm signal from the cosmic dawn. The relatively low cost, complexity, technology readiness level of electronics, and launch opportunities all make an Earth orbiting satellite an attractive and feasible option. Such a radiometer would perform better than a ground-based counterpart on two fronts, namely ionosphere induced chromaticity and systematics in antenna behavior from coupling to dielectrics in the near field. However, it will certainly receive RFI from Earth such as from FM transmitters in its field of view. Focusing only on FM based RFI, in this work we have investigated the presence of any low RFI zones in Earth orbit, the number of frequency channels that remain usable for signal detection after RFI flagging, and hence CD signal detection prospects in three Earth orbits - namely polar LEO, equatorial LEO, and GEO. We consider RFI only from FM transmitters on Earth, building a global database of the same, using free data from FMList.org and populating gaps in the data using a scalable statistical model. We find that there are indeed regions, such as around the Antarctic region and over the Pacific ocean where, within simulation resolution, there is no FM seeded RFI, assuming a wide antenna beam. While this is a lower limit based on FM seeded RFI alone, it is worth exploring this region for detecting CD signals in space around Earth. A polar LEO orbit maximizes time spent over this FM-radio-quiet zone and is the preferred orbit for maximizing signal detection sensitivity. Mock sky observations show that recovering CD/EoR signal over these FM radio quiet zones with high confidence is feasible for a majority of the standard theoretical models.
The work carried out in this paper only considers stationary FM based RFI sources for simulations. However, time varying RFI sources, including other satellites in Earth orbit pose a more complex problem to the experiment, and must be considered for future work carried out for CD/EoR science in Earth orbit. Temperature dependent effects on the properties of the antenna and any residual ionospheric effects (based on the altitude) are not considered in these simulations.\par
The FOM metric used in the simulations assume that there is no amplitude leakage from channels having very high RFI temperature. Accounting for this fact would decrease the percentage of CD/EoR models that can be detected with high confidence at high RFI regions. In all, we motivate the utility of a pilot experiment to map the RFI around Earth in space, particularly in polar orbit, to also capture the effect of other transient sources of RFI. This information of total RFI spectrum in space around Earth can pave the way for lower-cost and lower-complexity radiometer (than lunar farside) for CD detection radiometer orbiting Earth.

\section{Acknowledgments}
The authors would like to thank Sonia Ghosh for providing assistance with the STARFIRE algorithm. The authors would like to thank Vishakha Pandharpure for providing EM simulation data for the PRATUSH antenna. Y.P. extends special appreciation to Adarsh Kumar Dash and Tejas Oak for their helpful inputs during the project.

\appendix

\section{Tests for potential correlation of RFI with human activity}

\begin{figure}
  \centering
  \includegraphics[width=0.92\linewidth]{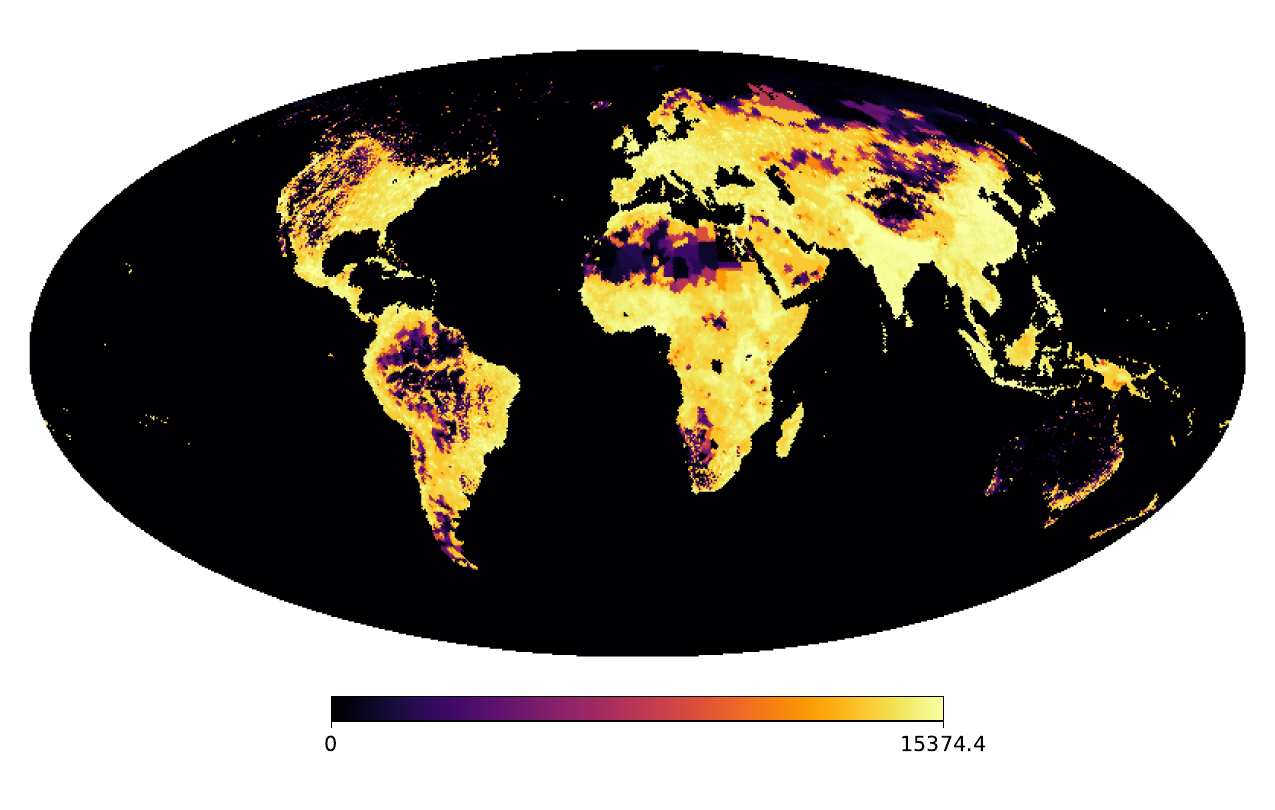}
  \caption{Global human population density HEALPix map of \NSide{} 128 }
  \label{fig:Pop dens map}
\end{figure}

Since the STARFIRE algorithm operates using the HEALPix pixellization scheme,  STARFIRE-2 aimed at conducting statistical studies using the same scheme. 
Population density was suspected to be correlated to the number of FM transmitters installed and hence the total RFI power in the FM band \citep{umar_RFI}. Hence studies were done to measure the extent of correlation between the two factors by comparing HEALPix heatmaps of population density with pixel-wise FM transmitter population and power. Obtaining a scalable statistical model mandates the usage of high-resolution data such as worldwide population density datasets available in the form of GEOTiff files. The pixels in a GEOTiff image are georeferenced \citep{mahammad2003geotiff}, indicating that they represent a scale affine transformation of the World Geodetic System-84 (WGS-84). One can confirm the accurate alignment of the right data points from the original image with the corresponding HEALPix pixel by ensuring that the defined spherical coordinates match the shape of the image. A 15 arc-minute resolution population density raster dataset was obtained from NASA-SEDAC\footnote{\href{https://sedac.ciesin.columbia.edu/data/set/gpw-v4-population-density-adjusted-to-2015-unwpp-country-totals-rev11/data-download }{SEDAC}} and processed into a HEALPix array using a python algorithm. \NSide{} 128 was chosen as the resolution of the HEALPix map throughout the correlation studies. (See Figure \ref{fig:Pop dens map})\par 

The STARFIRE-2 FM transmitter database contains information about various parameters of transmitters including their geodetic coordinates. This information was utilized to create an \NSide{} 128 HEALPix map of the transmitters on the globe with various parameters such as total transmitter power per pixel and number of transmitters per pixel.

As a preliminary test, the correlation studies were conducted for the limited countries' database present in the original STARFIRE database.
\par

  

\subsection{Using linear combinations of parameters for correlation studies}
Since direct comparison between population density and transmitter parameters seemed to be generally uncorrelated (Pearson's r=0.29), different factors such as Human Development Index (HDI) and country-wise land area seemed to be contributing factors for these parameters as well. Hence the independent variable of the correlation studies was modified into a linear combination of the above-mentioned factors and analyzed for potential trends. \par
A raster GEOTiff image of country-wise HDI map was created in QGIS (Quantum Geographic Information Systems) using a world-map shapefile. The shapefiles were connected to the HDI delimited text by utilizing the country name as the shared identifier between the two datasets. The resulting shapefile was rasterized with HDI feature as pixel weight and run through the same python algorithm thus creating an \NSide{} 128 HEALPix map of country-wise HDI. Hence the map was readily usable as a direct multiplicative factor with the population density HEALPix map for potentially better correlations since it contained exactly the same number of country-wise HDI weighted pixels as any other HEALPix map in use. However, there was no significant increase in Pearson's correlation coefficient (r=0.292). Hence it was concluded that the spatial distribution had no strong correlation with factors like population density but were rather randomly distributed with country specific transmitter power and frequency values.

\subsection{Correlation studies with complete database}
Further correlation studies were conducted using the newly complete database. Analyzing the trends between pixel-wise transmitter power and pixel-wise population density further solidified the notion that there was no strong correlation between these two parameters (Pearson's r=0.0099), but were rather randomly distributed with country dependent location and scale parameters.

\newpage
\bibliography{References}
\bibliographystyle{aasjournal}

\end{document}